\documentclass[aps,twocolumn,showpacs,eqsecnum]{revtex4}
\usepackage{epsfig}
\usepackage{amsbsy}
\usepackage{amsmath}
\usepackage{amsfonts}

\begin{document}
\title{Heterodyne and adaptive phase measurements on states of
fixed mean photon number}
\author{D.\ W.\ Berry, H.\ M.\ Wiseman, and Zhong-Xi Zhang} 
\affiliation{Centre for Laser Science, Department of Physics,
The University of Queensland, St.\ Lucia 4072, Australia}
\date{\today}

\begin{abstract}
The standard technique for measuring the phase of a single-mode field 
is heterodyne detection. Such a measurement may have an uncertainty 
far above the intrinsic quantum phase uncertainty of the state. Recently 
it has been shown [H.\ M.\ Wiseman and R.\ B.\ Killip, 
Phys.\ Rev.\ A {\bf 57}, 2169 (1998)] that an adaptive technique 
introduces far less excess noise. Here we quantify this difference by 
an exact numerical calculation of the minimum measured phase variance 
for the various schemes, optimized over states with a fixed mean 
photon number. We also analytically derive the asymptotics for these 
variances. For the case of heterodyne detection our results disagree
with the power law claimed by D'Ariano and Paris
[Phys.\ Rev.\ A {\bf 49}, 3022 (1994)].
\end{abstract}

\pacs{42.50.Dv, 03.67.Hk, 42.50.Lc}
\maketitle

\newcommand{\bra}[1]{\langle{#1}|} 
\newcommand{\ket}[1]{|{#1}\rangle}
\newcommand{\braket}[2]{\langle{#1}|{#2}\rangle}
\newcommand{\ip}[1]{\left\langle{#1}\right\rangle}
\newcommand{\nb}{\overline{n}}
\newcommand{\erfc}{\mathop{\rm erfc}}
\newcommand{\erf}[1]{Eq.\ (\ref{#1})}
\newcommand{\nn}{\nonumber} 

\section{Introduction}
\label{intro}
It is well known that it is not possible to make quantum-limited
measurements of the phase of an electromagnetic field using linear 
optics and photodetectors \cite{fullquan}. The standard
method of making a (non-quantum-limited) phase measurement is by heterodyne
detection. Heterodyne detection involves combining the field to be
measured with a much stronger local oscillator field which has a 
frequency detuned by a small amount $\Delta$, and measuring the
intensity of the
resultant field. In a typical experimental implementation, the
mode to be measured is passed through a 50/50 beam splitter, in order
to combine it with the local oscillator. The difference photocurrent
from the two output ports of the beam splitter yields a measurement of
the phase quadrature $X_\Phi = a e^{-i\Phi} + a^\dagger e^{i\Phi}$,
where $a$ is the annihilation operator of the mode to be measured and
$\Phi$ is the phase of the local oscillator. In heterodyne measurement
the detuning $\Delta$ is typically chosen to be large enough that 
the phase of the local oscillator cycles many times over the course of 
the measurement, so as to measure all quadratures with equal accuracy.

In a heterodyne phase measurement of a coherent state of amplitude 
$\alpha$, the variance in the measured phase $\phi$ scales as \cite{semiclass}
\begin{equation}
V_{\rm het}(\phi) = \frac{1}{2|\alpha|^{2}}.
\end{equation}
This is twice the
intrinsic uncertainty of the coherent state. An improved phase
measurement can be made if one has an initial estimate of the phase.
Then one would choose the local oscillator phase $\Phi$ to be equal to
$\varphi+\pi /2$, where $\varphi$ is the initial estimated phase of the field
to be measured. This is known as homodyne detection. If the phase is unknown
before the measurement, then one can still apply this idea by
adjusting the phase of the local oscillator 
during the course of the measurement based on an estimate of the phase
from the measurement results so far.

These adaptive phase measurements have been discussed in a series of
papers by Wiseman \cite{Wise95} and Wiseman and Killip 
\cite{semiclass,fullquan}. In Ref.\ \cite{semiclass} 
it was shown that via adaptive measurements of a
coherent state one can obtain a phase uncertainty of
\begin{equation}
V_{\rm adapt}(\phi)=\frac 1{4\alpha^2} + \frac 1{8\alpha^3}.
\end{equation}
Here we have taken the coherent amplitude $\alpha$ to be real.
The first term above is the intrinsic uncertainty of the coherent
state, and the second term is the extra phase uncertainty introduced
by the phase measurement. This result implies that even though there is not
much improvement in using adaptive phase measurements on coherent
states, if the input state has reduced phase uncertainty then adaptive
measurements will produce a phase measurement with far less
uncertainty than a heterodyne measurement.

To better quantify the improvement offered by adaptive measurements, 
it is necessary to consider the variance of states that have been optimized
for minimum phase variance under various measurement schemes. There must be a
constraint placed on optimizing states, otherwise a phase eigenstate will be
obtained. The two main ways in which to constrain the states are by truncating
the photon number and by fixing the mean photon number. The case of truncated
photon number was considered analytically in Ref.\ \cite{semiclass}, and
numerically in Ref.\ \cite{fullquan}. The case of a fixed mean photon number is
determined analytically in Sec.\ \ref{general} below.

General optimized states are difficult to work with, as they cannot be generated
experimentally. Also, in numerical integration of the stochastic differential
equations arising from the various measurement schemes (to be considered in
future work), calculations must be performed on the entire state for general
optimized states. This becomes prohibitively time consuming for large photon
numbers. Squeezed states are far more practical, as they are routinely generated
experimentally, and in numerical integration only the two squeezing parameters
need be considered. The theory for optimized squeezed states is considered in
Sec.\ \ref{squeezed}. The analytical results for both general states and
squeezed states are tested numerically in Sec.\ \ref{numeric}, and the
implications discussed in Sec.\ \ref{conclude}.

\section{POMs and phase measurements}
In quantum-mechanical systems, the most general way of obtaining the
probability of some measurement result $E$ is by the expectation value
of an operator $F(E)$, i.e.,
\begin{equation}
P(E)={\rm Tr}[\rho F(E)],
\end{equation}
where $\rho$ is the state matrix for the system. If the set of all possible
measurement results is $\Omega$, it is evident that $P(\Omega)=1$ for all
$\rho$, which implies that $F(\Omega)=1$. Thus $F(E)$ can be called a
probability operator, and the mapping $E \mapsto F$ defines a probability
operator measure (POM) on $\Omega$ \cite{Dav76,Hel76}.

For phase measurements on a single-mode field, the general form of the POM is
\cite{semiclass}
\begin{equation}
\label{POM}
F(\phi)=\frac{1}{2\pi} \sum_{n,m=0}^\infty \ket m \bra n e^{i\phi(m-n)}
H_{mn},
\end{equation}
where $H$ is a positive-semidefinite symmetric matrix with all entries
positive, and $\ket m$ is a number state of the field. In this case 
$\Omega = [0,2\pi)$ and the completeness relation is
\begin{equation}
\int_{0}^{2\pi} d\phi F(\phi) = {\rm 
diag}[H_{00},H_{11},H_{22},\ldots] = 1.
\end{equation}
For ideal phase measurements all elements of the $H$ matrix are equal to 1,
whereas for physical measurements the off-diagonal elements will generally be
less than 1 \cite{semiclass}.

The accuracy of phase measurements can be quantified in a number of ways
\cite{Hol84,Opatrny} which agree in the limit of small phase variance provided
the phase distribution is narrowly peaked (as it will be in the examples we
consider). When the mean phase is zero it is easiest to define the phase
variance as
\begin{equation}
\label{alternate}
V(\phi)=2-2\ip {\cos \phi} = 2(1-{\rm Re}\langle e^{i\phi}\rangle ).
\end{equation}
If we evaluate $\langle e^{i\phi}\rangle$ using Eq.\ (\ref{POM}) for an
arbitrary pure quantum state $\ket \psi$, we obtain
\begin{equation}
\langle{e^{i\phi}}\rangle =\sum_{m=0}^\infty \braket \psi m \braket {m+
1} \psi H_{m,m+1}.
\end{equation}
Therefore, the phase variance only depends on the off-diagonal elements
$H_{m,m+1}$, and we may characterize a phase measurement by the vector
\begin{equation}
h(m)=1-H_{m,m+1}.
\end{equation}
For most phase measurements, we have for large photon numbers 
\cite{semiclass}
\begin{equation}
h(m) \approx c m^{-p}.
\end{equation}
Then for states with a reasonably well-defined mean photon number 
$\bar{n}$ the total phase variance is given approximately by
\cite{semiclass}
\begin{equation}
V(\phi) \approx V_{\rm intrinsic}(\phi) + 2h(\bar{n}),
\end{equation}
where $V_{\rm intrinsic}(\phi)$ is the phase variance which would 
result from an ideal or canonical measurement of the phase
\cite{Lon27,LeoVacBohPau95}. Evidently $h(\bar{n})$ is a measure of the excess
phase noise introduced by the measurement, and it would be desirable to make
this as small as possible.

\section{Adaptive Measurements}
In a real optical experiment one cannot directly measure the phase, 
but would rather estimate it from a photocurrent record. Here the 
photocurrent would be derived by combining the mode to be
measured with a local oscillator via a 50/50 beam splitter. Such 
measurements have been called dyne measurements, as they include 
homodyne and heterodyne as well as adaptive measurements 
\cite{fullquan}. The signal
of interest is the difference between the photocurrents at the two
ports, which we define by
\begin{equation} \label{Ioft}
I(t) = \lim_{\delta t\rightarrow 0} \lim_{\beta\rightarrow\infty}
\frac {\delta N_+(t)-\delta N_-(t)}{\beta\delta t},
\end{equation}
where $\delta N_\pm$ are the increments in the 
photocounts at the two detectors in the interval $[t,t+\delta)$ and
$\beta$ is the amplitude of the local oscillator. 

Say the mode to be measured has an (assumed positive) envelope $u(t)$ normalized
such that $\int_{0}^{\infty}u(t)dt =1$. For simplicity, define a scaled time $v$
by
\begin{equation}
v=\int_0^t u(s) ds,
\end{equation}
so that $v \in [0,1)$.  
Then it turns out that at time $v$ there are two sufficient statistics
for the measurement record:
\begin{equation}
A_v=\int_0^v I(u) e^{i\Phi(u)}du, ~~~
B_v=-\int_0^v e^{2i\Phi(u)}du.
\end{equation}
These are sufficient statistics in the sense that the POM for the
measurement is a function of $A_v$ and $B_v$ only \cite{Wise96}.
This means that the best estimate for the phase at time $v$ 
need depend on the measurement record only through the complex numbers
$A_{v}$ and $B_{v}$.

Consider the simple case where the mode is initially 
in a coherent state $\ket{\alpha}$. We can then take the limits in 
\erf{Ioft} to obtain 
\begin{equation}
I(v) = 2\text{Re} (\alpha e^{-i\Phi(v)}) + \xi(v), 
\end{equation}
where $\xi(v)=dW(v)/dv$ is the shot noise \cite{semiclass}.
From this we can evaluate $A_v$ as
\begin{equation}
A_v=\alpha v -\alpha^* B_v+\int_0^v e^{2i\Phi(u)} dW(u).
\end{equation}
If one ignores the final term (which has zero expectation value),
it is easy to see that $\arg\alpha = \arg(v A_v+B_vA_v^*)$.
Other arguments \cite{Wise96} suggest that this is 
generally the best estimate for the phase at time $v$.
If $B_v$ is small (as it is in heterodyne detection), this can be
approximated by $\arg A_v$. 
The adaptive phase measurements that were analyzed in Refs.\ \cite{semiclass}
and \cite{fullquan} use $\arg A_v$ as the phase estimate during the measurement,
setting
\begin{equation}
\Phi(v) = \arg A_{v}+\pi/2.
\end{equation} The main motivations for this
choice are that it gives a feedback algorithm which would be easy to
implement experimentally, and that it is mathematically tractable.

If the phase estimate $\arg A$ is used at the end of the adaptive measurement,
the resultant phase measurement is actually worse than a heterodyne phase
measurement. This phase measurement is called an adaptive mark I phase
measurement, and it was found \cite{semiclass} that 
\begin{equation}
h_{\rm I}(m) \approx  \frac{1}{8m^{1/2}},
\end{equation}
which compares with 
\begin{equation}
h_{\rm het}(m) \approx \frac {1}{8 m}.
\end{equation}
If the phase estimate $\arg(A+BA^*)$ is used at the end of the phase
measurement, a far better phase measurement is obtained. This phase
measurement is called an adaptive mark II phase measurement, and 
yields \cite{semiclass}
\begin{equation}
h_{\rm II}(m) \approx \frac {1}{16 m^{3/2}},
\end{equation}
which is considerably better than the standard (heterodyne) result. However, as
noted in Sec.\ \ref{intro}, this dramatic improvement can only be seen if one
starts with a state having intrinsic phase variance much less than that of a
coherent state, for which $V_{\rm intrinsic}=2h_{\rm het}(\alpha^2)$.

In practice, imperfections in the equipment can introduce phase uncertainties
that scale as $\bar n^{-1}$ \cite{semiclass}. The main problems are
inefficient detectors and time delays. From Ref.\ \cite{semiclass}, the extra
phase variance introduced by inefficient detectors is
\begin{equation}
\delta V_{\rm II}(\phi)=\frac{1-\eta}{4\eta\nb},
\end{equation}
where $\eta$ is the efficiency. Again from Ref.\ \cite{semiclass} the extra
phase variance introduced by a time delay $\delta v$ is
\begin{equation}
\delta V_{\rm II}(\phi)\sim \frac{\delta v}{\bar n}.
\end{equation}
This means that when the photon number is sufficiently large, above
$(1-\eta)^{-2}$ or $(\delta v)^{-2}$, the introduced phase variance will always
scale as $\nb^{-1}$. Nevertheless, adaptive measurements will always give more
accurate results than heterodyne measurements, and it is of interest to know how
accurate phase measurements can be made in principle, since future technological
advances may greatly reduce the problems of detector inefficiencies and time
delays.

\section{General optimized states}
\label{general}
The fairest way to compare different phase measurement schemes is to consider
the phase variances for states optimized for minimum total phase variance for
the particular measurement under consideration. The two alternative constraints
which can be put on the optimization are (i) an upper limit on the photon
number, or (ii) a fixed mean photon number. The case where an upper limit $N$ is
put on the photon number was considered in Refs.\ \cite{semiclass,fullquan},
where the minimum phase variance was found to be
\begin{equation}
V(\phi)\approx 2cN^{-p}+(-z_1)(2cp)^{2/3}N^{-2(1+p)/3},
\end{equation}
where $z_1$ is the first zero of the Airy function, and is equal to
approximately $-2.338$.

Here we will be considering the case of fixed mean photon number.
Let us take the phase to be zero and use the operator for the phase
variance
\begin{subequations}
\begin{align}
\hat S &= 2-2\widehat{\cos{\phi}}\\
&=2- \sum_{m=0}^\infty \left[ 1-h(m) \right] \left( \ket m \bra
{m+1} + {\rm H.c.} \right).
\end{align}
\end{subequations}
In order to optimize for minimum phase variance with a fixed mean
photon number, we use the method of undetermined multipliers. We have
two constraints on the state: that the mean photon number is fixed and
that the state is normalized. This then gives us an equation with two
undetermined multipliers
\begin{equation}
\label{eigeq}
( \hat S + \mu \hat N ) \ket \psi = \nu \ket \psi.
\end{equation}
We solve this as an eigenvalue equation for $\nu$ with a fixed value
of $\mu$, and the eigenstate corresponding to the minimum eigenvalue
is the optimized state. The mean photon number can then be found from
the state. The mean photon number can be varied by varying $\mu$, but
cannot be easily predicted from $\mu$.

This method was used by Summy and Pegg \cite{meanN} to find states
with fixed mean photon number optimized for minimum intrinsic phase
variance. They found that the minimum phase uncertainty was
approximately
\begin{equation}
\ip {(\Delta\phi)^2}_{\rm intrinsic}^{\rm min}=\frac{C}{(\bar n + \epsilon)^2},
\end{equation}
where $C\approx 1.88$ and $\epsilon \approx 0.86$. Therefore, the
minimum intrinsic phase variance scales as $\bar n^{-2}$.

If the state is expressed in the number states basis
\begin{equation}
\ket \psi = \sum_{n=0}^\infty b_n \ket n,
\end{equation}
then in terms of $b_n$, \erf{eigeq} becomes
\begin{equation}
2b_n-\left[ 1-h(n) \right] \left( b_{n+1}+b_{n-1} \right) =
\left( \nu - \mu n \right) b_n.
\end{equation}
In the case of very large mean photon number, we can use 
a continuous approximation
\begin{equation}
2h(x)y-[ 1-h(x) ] \frac {d^2y}{dx^2} = \left(\nu-\mu
x\right) y,
\end{equation}
where we are taking $x=n$ and $y(x)=b_n$. For large $n$,
$\left[ 1-h(n) \right] \approx 1$, so
\begin{equation}
\label{deq1}
-\frac {d^2y}{dx^2}+y \left[2h(x)-\nu+\mu x\right] \approx 0.
\end{equation}
Now define
\begin{equation}
f(x)=2h(x)-\nu+\mu x.
\end{equation}
Then, expanding $f(x)$ in a Taylor series around 
\begin{equation}
x_0=(\mu/2cp) ^{-1/(p+1)} ,
\end{equation}
we find
\begin{subequations}
\begin{align}
f(x_0)&=2h(x_0)-\nu+\mu x_0, \\
f'(x_0)&=2h'(x_0)+\mu=-2cpx_0^{-p-1}+\mu=0, \\
 f''(x_0)&=2h''(x_0)=2cp(p+1)x_0^{-p-2} ,\\
 f'''(x_0)&=-2cp(p+1)(p+2)x_0^{-p-3}.
\end{align}
\end{subequations}
This technique requires that the number distribution has its
maximum near $x_0$, which will be justified below.

Using the Taylor series for $f(x)$, and defining $f_{0}=f(x_{0})$, 
$f_{2} = f''(x_{0})/2$, and $f_{3} = f'''(x_{0})/6$, the differential 
equation (\ref{deq1}) becomes
\begin{equation}
\label{eigeq2}
-\frac {d^2y}{dx^2}+ [ f_{2}(x-x_0)^2+{f_{3}}(x-x_0)^3]y \approx -f_0y.
\end{equation}
Note that $-f_0=\nu - [2h(x_0)+\mu x_0]$, so the above equation is
equivalent to solving \erf{eigeq} as an eigenvalue equation for $\nu$
with a fixed value of $\mu$. Now \erf{eigeq2} is equivalent to the 
time-independent Schr\"odinger equation with energy eigenvalue
$E=-f_{0}$ for a perturbed harmonic Hamiltonian $\hat{H}=\hat H_0+\hat H_1$,
where
\begin{subequations}
\begin{align}
\hat H_0&=\sqrt{f_2}\left[-\frac {d^2}{d\xi^2}+ \xi^2\right],\\
\hat H_1&=b\xi^3,
\end{align}
\end{subequations}
where
\begin{subequations}
\begin{align}
b&=-\frac{p+2}3 \left[cp(p+1)\right]^{1/4}x_0^{-p/4-3/2},\\
\xi&=f_{2}^{1/4}(x-x_0).
\end{align}
\end{subequations}

The unperturbed solution is
\begin{equation}
y_j^{(0)}(\xi)=(\pi^{1/2}2^jj!)^{-1/2} \exp(-\xi^2/2)H_j(\xi),
\end{equation}
where $H_j$ are Hermite polynomials. This solution is only valid for 
$\sqrt{f_{2}}x_{0}^{2} \gg 1$, which requires that $p<2$
in addition to $x_{0}\gg 1$. The energy eigenvalues are
\begin{equation}
E_j^{(0)}=(2j+1)\sqrt{f_{2}}.
\end{equation}
The lowest energy eigenvalue $E_0$ and corresponding eigenstate $y_0$ will
minimize $\nu$ and therefore minimize $\langle \hat S\rangle$. From perturbation
theory they can be expressed as
\begin{subequations}
\begin{align}
E_0 &\approx E_0^{(0)}+\bra{0}\hat H_1\ket{0}-
\frac{b^2}{2\sqrt{f_{2}}}
\sum_{k=1}^\infty\frac{\left|\bra{k}\xi^3\ket{0}\right|^2}k , \\
y_0 &\approx y_0^{(0)}-\frac {b}{2\sqrt{f_{2}}}
\sum_{k=1}^\infty\frac{\bra{k}\xi^3\ket{0}}k y_k^{(0)},
\end{align}
\end{subequations}
where $\ket{j}$ is the state corresponding to $y_j^{(0)}(\xi)$.
Now it is easily shown using the properties of Hermite polynomials
that the only nonzero terms are $\bra{1}\xi^3\ket{0}=3/{2\sqrt 2}$ and 
$\bra{3}\xi^3\ket{0}={\sqrt 3}/2$.
This then gives the lowest energy eigenvalue and eigenstate as
\begin{subequations}
\begin{align}
E_0 & \approx E_0^{(0)}-b^2 \frac{11}{16}f_{2}^{-1/2} , \\
y_0 & \approx y_0^{(0)}-\frac {b}{4\sqrt{2f_{2}}}
\left[3y_1^{(0)}+\sqrt\frac23 y_3^{(0)}\right].
\end{align}
\end{subequations}

Now we can use these expressions to find the mean photon number as
\begin{subequations}
\begin{align}
\bar n&=\ip{x}=\int y_0(\xi)(x_0+f_2^{-1/4}\xi)y_0(\xi)d\xi \\
&\approx x_0 + \frac{p+2}{4\sqrt{cp(p+1)}}x_0^{p/2}.
\end{align}
\end{subequations}
As we are assuming $p<2$,
\begin{equation}
x_0 \approx \bar n\left[1-\frac{p+2}{4\sqrt{cp(p+1)}}\bar n^{p/2-1}\right],
\end{equation}
so that the mean photon number is close to $x_0$, justifying the previous
expansion around $x_0$. Now we can find the minimum phase variance using
\begin{equation}
\langle\hat{S}\rangle_{\rm min} = (\nu -\mu \bar n )_{\rm min},
\end{equation}
and $\nu=-f_0+2h(x_0)+\mu x_0$, $\mu=2cpx_0^{-p-1}$, and $f_0=-E_0$. We obtain
the first two terms as
\begin{equation} \label{finalgen}
\langle\hat{S}\rangle_{\rm min} \approx 2c\bar n^{-p}+\sqrt{cp(p+1)}\bar
n^{-p/2-1}.
\end{equation}
Note that the first term here is the same as the result when an upper
limit is put on the photon number, but the second term scales as
a different power of $\bar n$.

A particular case of interest is heterodyne detection, for which we find 
\begin{equation}
\langle\hat{S}\rangle^{\rm min}_{\rm het} \approx \frac{1}{4\bar{n}}
+ \frac{1}{2\bar{n}^{3/2}}.
\end{equation}
This is interesting because it differs radically from the result 
claimed by D'Ariano and Paris \cite{DArPar94} of
\begin{equation} \label{Dar}
\langle(\Delta \phi)^{2}\rangle_{\rm het}^{\rm min} = 
\frac{1.00\pm 0.02}{\bar{n}^{1.30\pm 0.02}}.
\end{equation}
As the quoted errors suggest, this result was obtained entirely 
numerically, in contrast to our analytical result. In Sec.\ \ref{numeric} we 
present our own numerical results and show that our analytical result 
is a far better fit than the power law of D'Ariano and Paris. 

\section{Optimized squeezed states}
\label{squeezed}
As an alternative to considering general optimized states, we can
consider optimized squeezed states. There are three reasons for
this:

(1) Squeezed states are relatively easily generated in the laboratory, whereas
there is no known way of producing general optimized states
experimentally.

(2) Squeezed states can be treated numerically far more easily than
general optimized states.

(3) It has been found numerically (see Sec.\ \ref{numeric}) that the
phase uncertainties of optimized squeezed states are very close to
those of optimized general states, and a partial theoretical
explanation can be obtained by the following analysis.

Squeezed states optimized for minimum intrinsic phase variance were
previously considered by Collett \cite{Collett93}, who found that
the minimum phase uncertainty was approximately given by
\begin{equation}
\langle(\Delta\phi)^2\rangle_{\rm intrinsic} 
\approx \frac {\ln(\bar n)+\Delta}{4\bar n^2},
\end{equation}
where $\Delta=\frac 32 +2\ln 2-\frac 14 \ln(2 \pi)$. The scaling
of optimized squeezed states is therefore worse than the scaling of
optimized general states when we consider intrinsic phase variance.
The difference is only a factor of $\ln \bar n$, however.

We consider squeezed states of the form
\begin{equation}
\ket{\alpha,\zeta}=\exp(\alpha a^\dagger -\alpha^* a)\exp[(\zeta^*
a^2-\zeta a^{\dagger 2})/2]\ket 0.
\end{equation}
Now we will take the phase to be 0, so $\alpha$ is real. Phase squeezed states
have $\zeta$ real and negative, so we can take $\zeta$ to be real. Then using
the definition of phase uncertainty in \erf{alternate} gives 
\begin{align}
\langle(\Delta\phi)^2\rangle&=2-2\sum_{n=0}^\infty \braket{\alpha,\zeta}{n}
\braket{n+1}{\alpha,\zeta}\nn \\
& ~~~ +2\sum_{n=0}^\infty h(n)\braket{\alpha,\zeta
}{n}\braket{n+1}{\alpha,\zeta}. \label{3term}
\end{align}
In estimating the intrinsic phase uncertainty, Collett
\cite{Collett93} found the first two terms on the right-hand side to be
approximately $(n_0+1)/4\nb^2+2\erfc(\sqrt{2n_0})$, where $n_0=\nb e^{2\zeta}$.
We therefore only have to determine an expression for the third term. Using
$h(n) \approx cn^{-p}$ as usual, the result we require is derived in the
Appendix:
\begin{equation} \label{appresult}
\sum_{n=0}^\infty n^{-p} \braket{\alpha,\zeta}{n}\braket{n+1}
{\alpha,\zeta} \approx \nb^{-p}\left[ 1+\frac {p(p+1)}{2n_0} \right].
\end{equation}
The phase uncertainty is therefore given by
\begin{align}
\langle(\Delta\phi)^2\rangle \approx \frac{n_0\!+\!1}{4\nb^2} + 2
\erfc(\sqrt{2n_0}) + 2c\nb^{-p} \!\left[ 1\!+\!\frac {p(p+1)}{2n_0}
\right]\!. \nn \\ \label{phunc1}
\end{align}
Taking the derivative with respect to $n_0$ gives
\begin{equation}
\frac {\partial\langle(\Delta\phi)^2\rangle}{\partial n_0}\approx\frac 1{4\nb^2}
-\sqrt{\frac 8{\pi n_0}}e^{-2n_0}+2c\nb^{-p}\left[\frac{-p(p+1)}{2n_0^2}\right].
\end{equation}
As the second term falls exponentially with $n_0$, it can be omitted.
Then we find that the minimum phase variance occurs for
\begin{equation}
\label{minn0}
n_0 \approx 2\sqrt{cp(p+1)}\nb^{1-p/2}.
\end{equation}
Thus we find that $n_0 \propto \bar n^{1-p/2}$, as was used in the Appendix.
Substituting this result into \erf{phunc1} gives
\begin{equation}
\langle(\Delta\phi)^2\rangle_{\rm min}\approx 2c\nb^{-p}+\sqrt{cp(p+1)}
\nb^{-p/2-1}.
\end{equation}
We therefore obtain exactly the same first two terms for the phase
uncertainty when considering squeezed states as we do when considering
general states.

\section{Numerical results}
\label{numeric}
The analytic results from Secs.\ \ref{general} and \ref{squeezed} have been
verified numerically by calculating the optimized states for heterodyne
measurements and adaptive mark I and II measurements. For moderate mean photon
numbers the calculations were exact, except in that a cutoff at large photon
numbers was used.

For larger photon numbers an additional approximation was that asymptotic
expressions for $h(m)$ were used. For heterodyne measurements there is an
exact expression for $h(m)$ \cite{fullquan},
\begin{equation}
h_{\rm het}(m)=1-\frac{\Gamma\left(m+3/2\right)}{\sqrt{\Gamma\left(
m+1\right)\Gamma\left(m+2\right)}}.
\end{equation}
This form of the equation is very inconvenient for use in numerical work due to
roundoff error. It is more convenient to use the asymptotic expansion. From the
asymptotic expansion of $\ln \Gamma(m)$ \cite{Erdelyi53} it is simple to find
an expansion of $h_{\rm het}(m)$ in powers of $1/(m+1)$ to as many terms as
required. The first few terms are given by
\begin{equation}
h_{\rm het}(m) = \frac{1}{8 (m+1)}-\frac{1}{2^7(m+1)^{2}}+
O\left(m^{-3}\right).
\end{equation}
An expansion up to 12th order was used to determine $h_{\rm het}(m)$. This
expansion was found to be more accurate than using the formula directly for
values of $m{\gtrsim}12$.

For adaptive mark I and adaptive mark II measurements, $h(m)$ can be
determined using methods discussed in Ref.\ \cite{fullquan}. For the mark I
case $h_{\rm I}(m)$ was determined exactly up to $m=3000$. Further
values were extrapolated by fitting an asymptotic expansion to the
results below 3000. The first three terms of the asymptotic expansion
used were
\begin{equation} \label{adap1}
h_{\rm I}(m) = \frac{1}{8 
m^{1/2}}-\frac{0.101562}{m}-\frac{0.0508}{m^{3/2}} 
+ O(m^{-2}).
\end{equation}
The uncertainties in these numbers were estimated using various methods of
fitting, and are indicated by the number of significant figures quoted.

For the mark II case $h_{\rm II}(m)$ was determined
exactly up to $m=1000$. Fitting techniques did not consistently
give any higher-order terms than that obtained by the semiclassical
theory, so for $m>1000$ the formula $h_{\rm II}(m)=\frac 1{16}m^{-3/2}$
was used.

For very large mean photon numbers ${\gtrsim}10^5$, it was not
feasible to solve the exact eigenvalue problem, but an approximate
solution was obtained by using the continuous approximation of the
eigenvalue problem and discretizing it. In order to reduce the number
of intervals required in the discretized equation, the equation was
solved for three different numbers of intervals.
The result for the continuous case was then estimated by
projecting to zero step size, assuming the error is quadratic in the
step size. For optimized squeezed states the full calculation was
performed up to a mean photon number of about $10^6$, beyond which
roundoff error became too severe.

\begin{figure}
\includegraphics[width=0.45\textwidth]{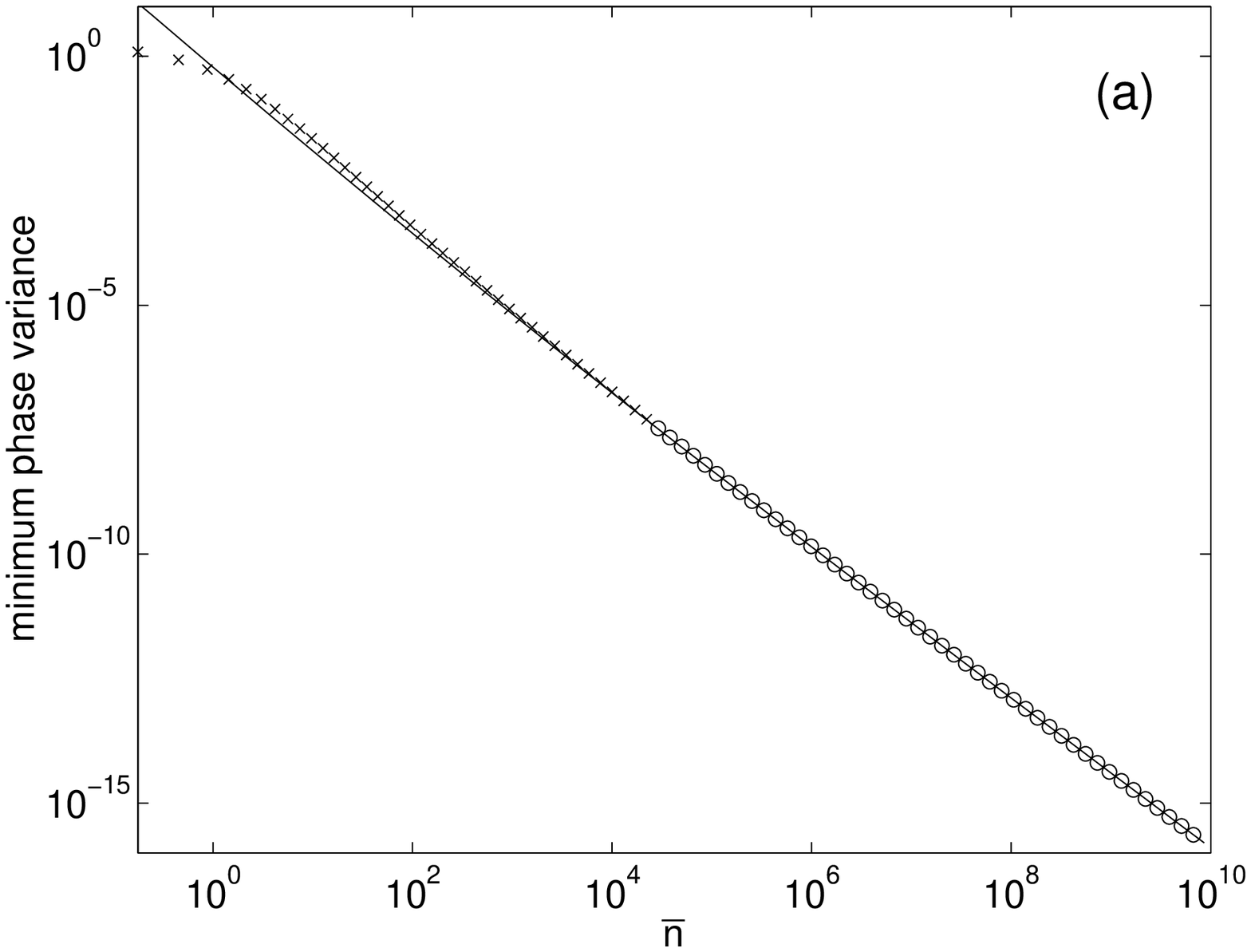}
\includegraphics[width=0.45\textwidth]{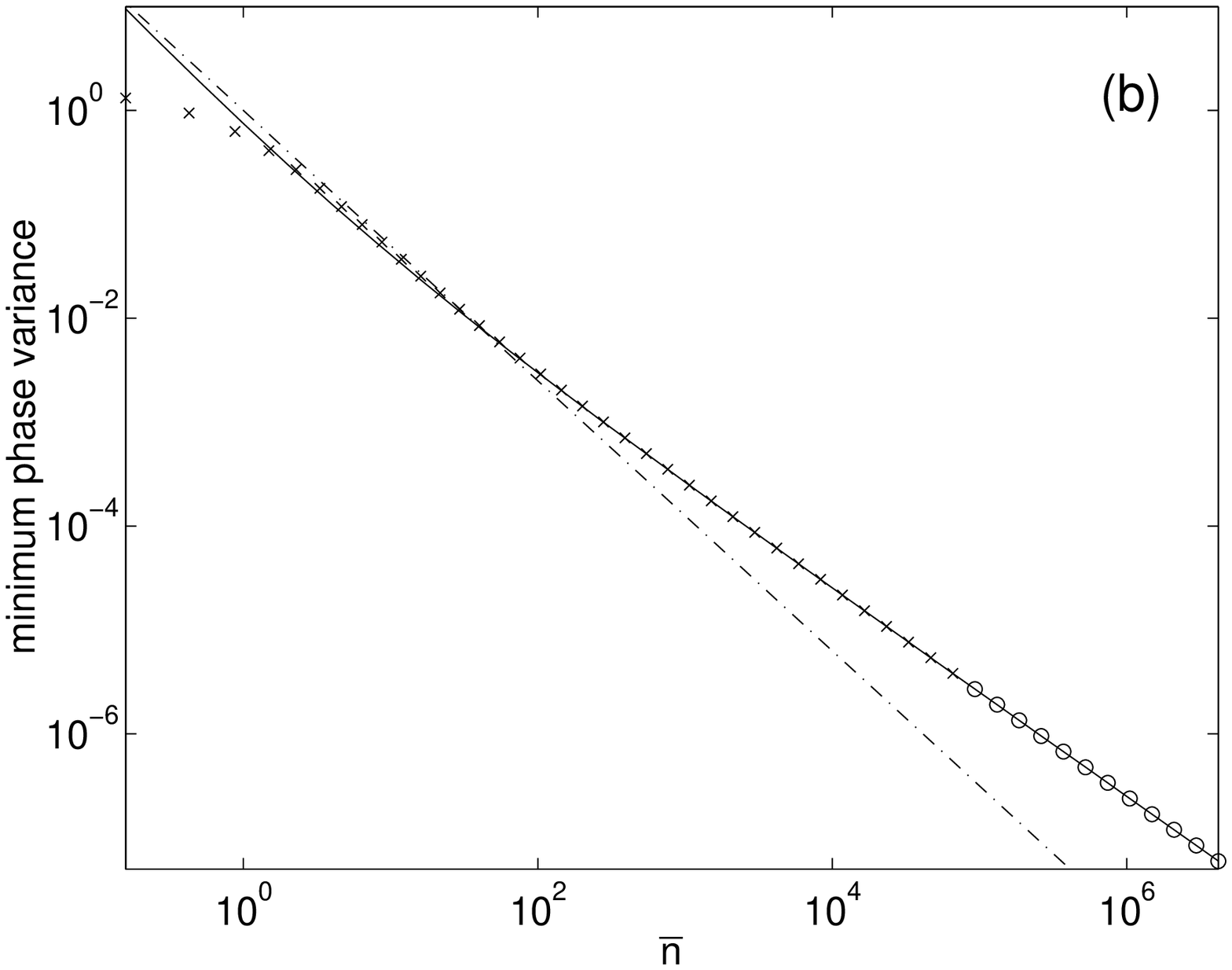}
\includegraphics[width=0.45\textwidth]{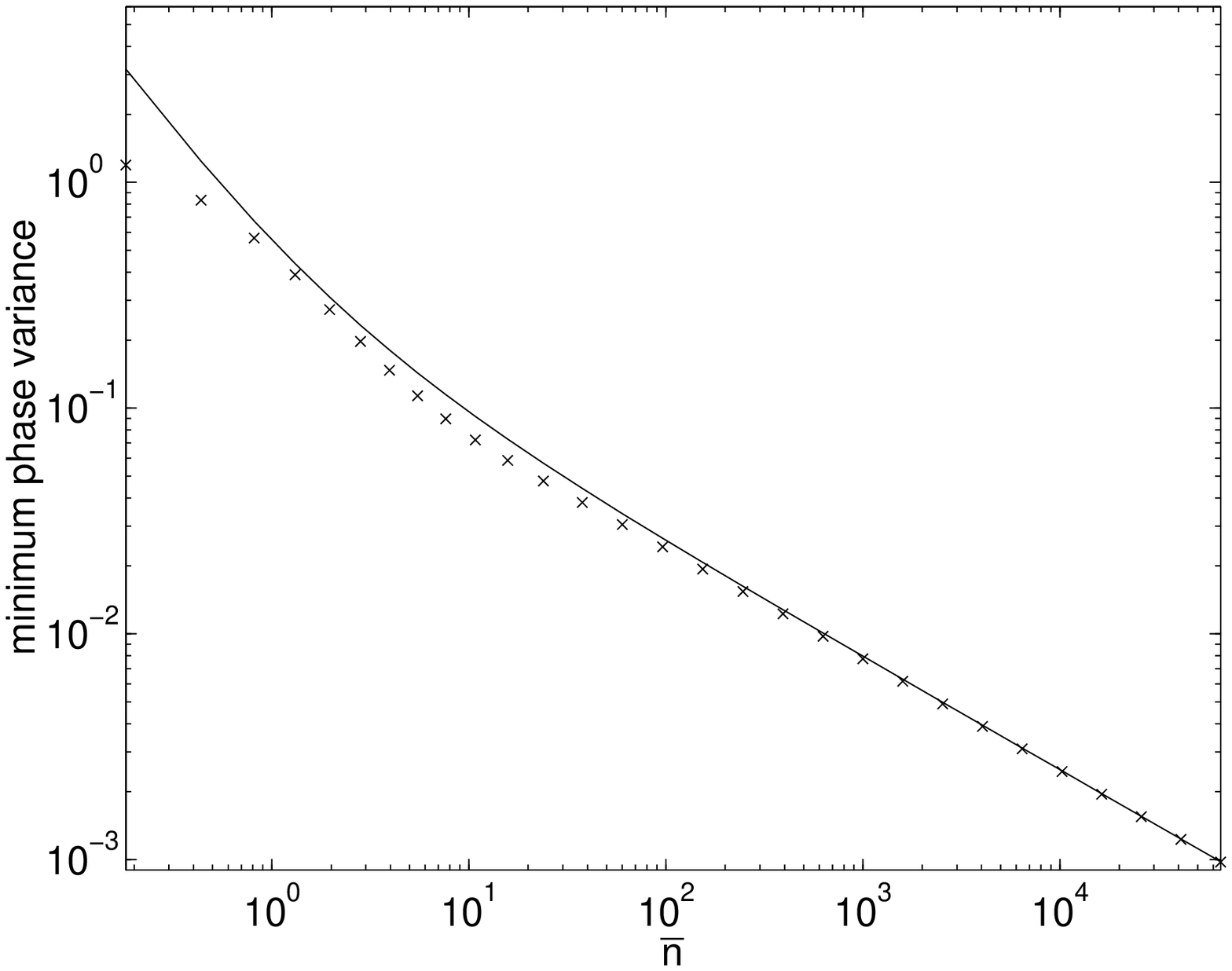}
\caption{Plot of the phase variance for general optimized states via complete
eigenvalue solution (crosses) and continuous approximation (circles) vs mean
photon number $\bar{n}$. The asymptotic analytical expression from Sec.\ 
\ref{general} is also plotted (continuous lines). The three phase detection
schemes are (a) mark II adaptive, (b) heterodyne, and (c) mark I adaptive. The
power law claimed by D'Ariano and Paris for heterodyne detection is also plotted
in (b) (dash-dot line). }
\label{loglog}
\end{figure}

The results of the numerical calculation for the general optimized states are
shown in Fig.\ \ref{loglog}, along with the analytical results obtained in
Sec.\ \ref{general} and the power law of \erf{Dar} published by D'Ariano and
Paris \cite{DArPar94}. It is seen that the results for heterodyne measurements
agree reasonably well with the power law of D'Ariano and Paris for moderate
photon numbers (up to about 100). Above this, however, the agreement is
extremely poor. This is presumably due to the fact that the numerical data used
by D'Ariano and Paris seems to have been limited to maximum photon numbers only
of order 100. In contrast, our analytical result agrees very well for mean
photon numbers above about 100. The analytical result also agrees well with the
numerical results for the adaptive mark I and II measurements.

\begin{figure}
\includegraphics[width=0.45\textwidth]{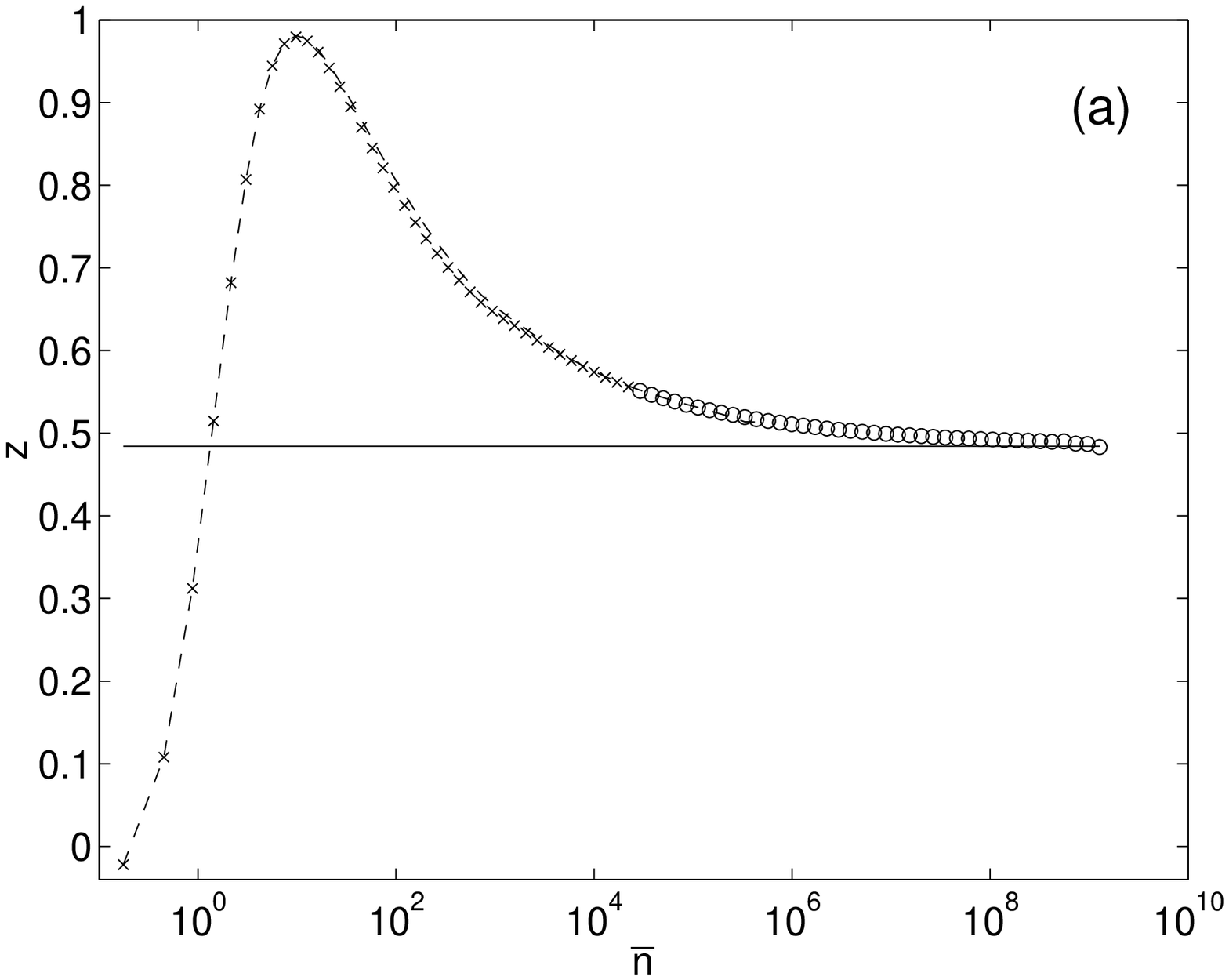}
\includegraphics[width=0.45\textwidth]{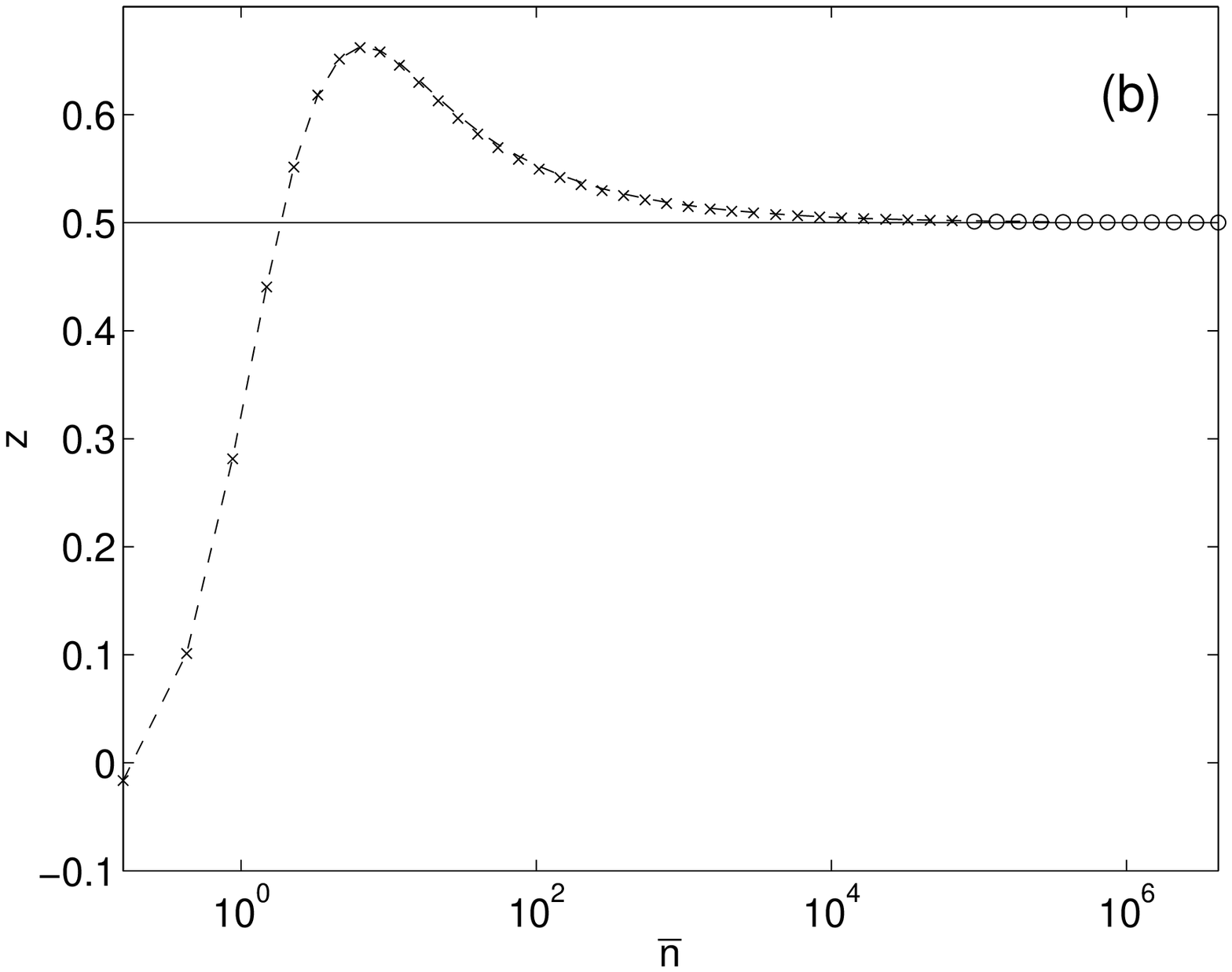}
\includegraphics[width=0.45\textwidth]{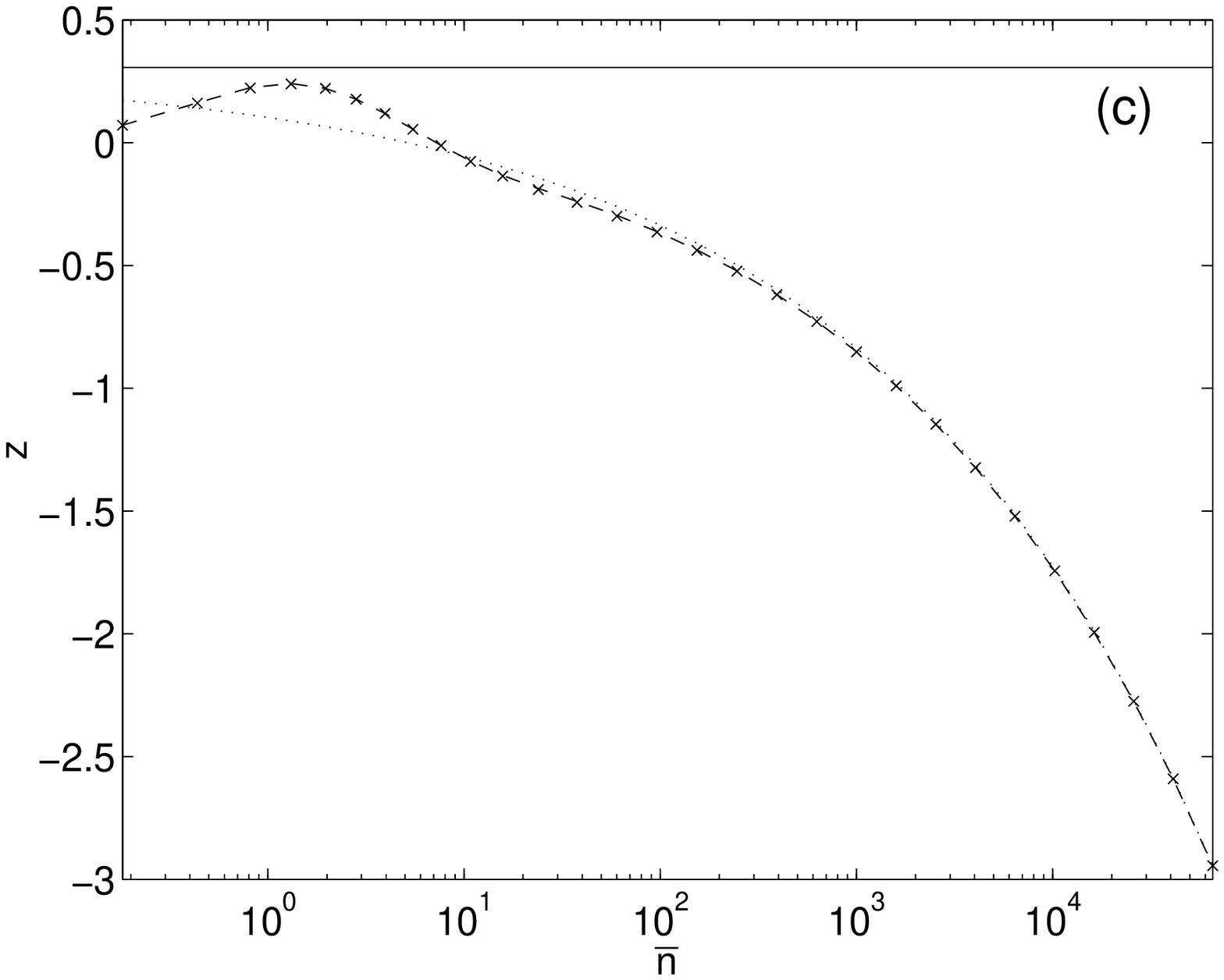}
\caption{Plot of $z$ parameter from \erf{defz} for general optimized states via
the complete eigenvalue solution (crosses) and continuous approximation
(circles), and optimized squeezed states (dashed lines), vs mean photon number
$\bar{n}$. The theoretical asymptotic value of $\sqrt{cp(p+1)}$ is shown as the
continuous horizontal lines. The results shown are for mark II adaptive phase
measurements in (a), heterodyne measurements in (b), and mark I adaptive
measurements in (c). The predicted $z$ for the mark I case taking into account
the second term for $h_{\rm I}(m)$ is also plotted in (c) (dotted line).}
\label{calculations}
\end{figure}

On the log-log plot of the phase variance it is extremely difficult to
discriminate between different phase variances unless the difference
is greater than a factor of 2. Therefore, in Fig.\ \ref{calculations} we plot
the parameter $z$, defined by
\begin{equation} \label{defz}
z=(\langle(\Delta \phi)^2\rangle_{\rm min}-2c\bar n^{-p})\bar n^{p/2+1}.
\end{equation}
From the above analysis this parameter should converge to $\sqrt{cp
(p+1)}$ for large $\bar n$. The $z$ parameter is plotted for optimized
general states and optimized squeezed states for the cases of mark
II, heterodyne and mark I measurements in Figs.\ \ref{calculations}(a),
\ref{calculations}(b), and \ref{calculations}(c), respectively.

In the adaptive mark II and heterodyne results in Fig.\ \ref{calculations}(a)
and \ref{calculations}(b), we can see that the results using the full eigenvalue
solution match up very well with the continuous approximation results,
demonstrating the accuracy of this technique. In addition, the squeezed-state
results are extremely close to the general optimized state results, far closer
than indicated by the first two terms derived above. Also note that in
Fig.\ \ref{calculations}(a) the results do not agree closely (within 1\%) with
the asymptotic value until $\bar{n}>10^8$, whereas the heterodyne results
converge at a much lower photon number, with good agreement for $\nb{\approx}
10^4$.

In the adaptive mark I results in Fig.\ \ref{calculations}(c), there is again
good agreement between the squeezed-state results and the general optimized
state results. However these results do not approach the asymptotic value at all
in this case. The reason for this is that there is a higher-order term in
$h_{\rm I}(m)$ which is of order $m^{-1}$, as shown in \erf{adap1}. This term is
of lower order than the second term in the expansion (\ref{finalgen}) of the
phase variance, which is of order $m^{-5/4}$. If the higher-order term in
$h_{\rm I}(m)$ is taken into account in predicting the asymptotic value, then
there is good agreement with the numerical results.

Although the adaptive mark I scheme has the poorest performance for large mean
photon numbers, note from Fig.\ \ref{loglog} that for small mean photon
numbers (of order unity) it is actually the best scheme, having the 
smallest minimum phase variance. This is to be expected from the 
results of previous work \cite{Wise95,fullquan}, showing that for 
a {\em maximum} photon number of 1, the adaptive mark I scheme is 
actually the best possible.

The significance of these results is, first, that the two terms
given by the theory for optimized general and squeezed states are
correct, and, second, that the phase uncertainties of optimized
squeezed states are extremely close to those for optimized general
states. This result is of great importance, as it means that in
numerical and experimental work squeezed states can be used rather
than general states.

This is an advantage in numerical work because only the squeezing
parameters need be considered, rather than the entire state. This
means that, for example, a numerical evaluation of different feedback
schemes is feasible, and this work is being carried out now. In
experimental work it is an advantage because squeezed states can be
produced experimentally, whereas arbitrary states cannot. This means
that it is possible to produce states experimentally that are very
close to optimized for the different measurement schemes.

\section{conclusion}
\label{conclude}
We have derived asymptotic analytical expressions for the minimum phase 
variances obtainable under various detection schemes, for states 
constrained  by their mean photon number. The detection schemes 
considered were heterodyne detection (a standard scheme), and two 
single-shot adaptive schemes (first proposed in Refs.\ \cite{Wise95,semiclass}),
called mark I and mark II. Numerical results confirm 
the correctness of the first two terms in the asymptotic expansion, 
except in one case (adaptive mark I), where the second term was not 
expected to be correct. Furthermore, analytical and numerical results 
show that essentially the same results may be obtained using squeezed 
states, rather than completely general states. This is an important 
result from both theoretical and experimental standpoints.

As expected, the minimum phase variance for adaptive mark II 
measurements was much smaller than that for the standard technique of 
heterodyne detection, for large mean photon number $\nb$. In 
particular, the leading term in the former scaled as $\nb^{-3/2}$ 
compared to $\nb^{-1}$ in the latter. The claim by D'Ariano and Paris 
\cite{DArPar94} 
that the heterodyne phase variance scaled as $\nb^{-1.30}$ 
(or, as stated in their abstract, $\nb^{-4/3}$) 
 was proven wrong. This reinforces the position of 
adaptive mark II phase measurements as the best known phase 
measurement scheme.

In Ref.\ \cite{fullquan} it was shown that a lower bound for the phase
uncertainty introduced by measurements is $(\ln\nb)/(4\nb^2)$.
For large photon numbers this is a lot smaller than the phase variance
of $1/(8\nb^{3/2})$ introduced by mark II measurements. 
It therefore may be possible to obtain a higher power in the scaling
law by modifying the measurement method. The most promising modification
is using a different feedback phase, and this is currently under
investigation numerically via the solution of stochastic Schr\"odinger 
equations \cite{Car93b,Wise96}. This is possible even with very large 
photon numbers if one uses squeezed states, because these remain 
squeezed states even under the stochastic evolution. The results 
obtained in this paper justify this approach, as the variances 
obtained for the cases of general states and squeezed states 
were almost indistinguishable.
 
\acknowledgments
This work was supported by the Australian Research Council and the 
University of Queensland.
	 
\appendix
\section*{Appendix: Deriving Eq.\ (\ref{appresult})}
\renewcommand{\theequation}{A\arabic{equation}}

We wish to evaluate the following sum:
\begin{equation} \label{A1}
\sum_{n=0}^\infty h(n) \braket{\alpha,\zeta}{n}\braket{n+1}
{\alpha,\zeta}.
\end{equation}
We can do this in the following way. The number state representation
of squeezed states is given by \cite{Yuen76}
\begin{align}
\braket{n}{\alpha,\zeta} &= (n!\mu)^{-1/2}(\nu/2\mu)^{n/2}H_n[\beta(2\mu
\nu)^{-1/2}] \nn \\ & ~~~ \times \exp[-|\beta|^2/2+(\nu^{*}/2\mu)\beta^2],
\end{align}
where
\begin{equation}
\mu=\cosh r, ~~~ \nu=e^{i\phi} \sinh r, ~~~ \beta=\alpha \mu + \alpha^* \nu.
\end{equation}
Here $r$ and $\phi$ are the magnitude and phase, respectively, of
$\zeta$, while $H_n$ are Hermite polynomials and satisfy the recursion
relation \cite{Erdelyi53}
\begin{equation}
\label{Recurse}
H_{n+1}(x)-2xH_n(x)+2nH_{n-1}(x)=0.
\end{equation}
This means that the number representation of squeezed states satisfies
the recursion relation
\begin{equation}
\braket{n+1}{\alpha,\zeta}\mu\sqrt{n+1}-\braket{n}{\alpha,\zeta}\beta
+\braket{n-1}{\alpha,\zeta}\nu\sqrt{n}=0.
\end{equation}
Rearranging this and squaring gives
\begin{align}
&|\braket{n+1}{\alpha,\zeta}|^2\mu^2(n+1)=|\braket{n}{\alpha,\zeta}|^2\beta^2
\nn \\ & ~~~ +|\braket{n-1}{\alpha,\zeta}|^2\nu^2n-2\braket{\alpha,\zeta}{n}
\braket{n-1}{\alpha,\zeta}\beta\nu\sqrt{n}.
\end{align}
This expression is only true for real squeezing parameters.
Multiplying this by $n^k$ and summing gives
\begin{align}
2\beta\nu &\sum_{n=0}^\infty (n+1)^{k+1/2}
\braket{\alpha,\zeta}{n}\braket{n+1}{\alpha,\zeta} \nn \\ &=\beta^2\langle{n^k}
\rangle+\nu^2\langle{(n+1)^{k+1}}\rangle-\mu^2\langle{n(n-1)^k}\rangle.
\end{align}
Now let us take $-p=k+\frac 12$. In this case we find that the shift of
indices cannot be performed exactly, but the contribution from terms
near $n=0$ will be negligible. Also some of the terms above diverge
near $n=0$; however, the divergent terms are the extra terms produced
by the shift of indices, and in the following expansions the behavior
near $n=0$ is ignored.

Taking $-p=k+\frac 12$ and considering the deviation from the mean photon
number gives
\begin{align}
2\beta & \nu \sum_{n=0}^\infty (n+1)^{-p}\braket{\beta}{n}\braket{n+1}{
\beta} \nn \\ &\approx\beta^2\big\langle(\nb+\Delta n)^{-(p+1/2)}\big\rangle
+\nu^2\big\langle(\overline{n}+1+\Delta n)^{-(p-1/2)}\big\rangle \nn \\
&~~-\mu^2\big\langle (\overline{n}-1+\Delta n)^{-(p-1/2)}\big\rangle \nn \\ &
~~-\mu^2\big\langle(\nb-1+\Delta n)^{-(p+1/2)}\big\rangle .
\end{align}
Expanding this in a series in $\Delta n$ gives
\begin{align}
\label{series}
2\beta\nu &\sum_{n=0}^\infty (n+1)^{-p}
\braket{\beta}{n}\braket{n+1}{\beta} \nn \\
&\approx \sum_{j=0}^\infty \frac {(-1)^j\langle\Delta n^j\rangle}{j!}\left\{
\frac{(p+j-1/2)!}{(p-1/2)!}(
\beta^2\nb^{-(p+j+1/2)} \right. \nn \\
& ~~~~~-\mu^2(\nb-1)^{-(p+j+1/2)})+\frac{(p+j-3/2)!}{(p-3/2 )!}
\nn \\ & ~~~~~\left. \times [\nu^2 (\nb+1)^{-(p+j-1/2)}
-\mu^2(\nb-1)^{-(p+j-1/2)}] \vphantom{\frac{(p3/2)!}{(p3/2)!}} \right\}. \nn \\
\end{align}

Now we have an expression we can use to evaluate \erf{A1}. Recall that for
generalized measurements we have the asymptotic expression
$h(n){\approx}cn^{-p}$. This is equivalent to $h(n)\approx c (n+1)^{-p}$, as the
difference is of higher order. It is easily shown that for squeezed states
$\langle\Delta n^2\rangle=\alpha^2(\mu-\nu)^2+2\mu^2\nu^2$. Therefore, using the
first three terms of \erf{series} above, we find
\begin{align} 
2\beta\nu &\sum_{n=0}^\infty (n+1)^{-p}\braket{\alpha,\zeta}{n}\braket
{n+1}{\alpha,\zeta} \nn \\ & \approx \beta^2\nb^{-(p+1/2)}-\mu^2(\nb-1)^
{-(p+1/2)} \nn \\ &~~+\nu^2(\nb+1)^{-(p-1/2)}-\mu^2(\nb-1)^{-(p-1/2)} \nn \\
& ~~+\!\left[\frac {\alpha^2(\mu-\nu)^2}{2}+\mu^2\nu^2\right]
\left\{ \left(p^{2}+2p+\frac 34\right)\right. \nn \\ &~~\times (
\beta^2 \nb^{-(p+5/2)}-\mu^2(\nb-1)^{-(p+5/2)})+\left(p^{2}-\frac 14 \right)\nn
\\ & ~~\left. \times [\nu^2(\nb+1)^{-(p+3/2)}-\mu^2(\nb-1)^{-(p+3/2)}]
\vphantom{\frac 34}\right\}\!. \nn \\ \label{first3}
\end{align}
At this stage the main problem is to determine which terms should be
kept. This depends on how $n_0$ scales with $\bar n$. If the state is
optimized for minimum intrinsic phase uncertainty, then
$n_0\propto\ln(\nb)$ \cite{Collett93}. If we carry the derivation
through using this result to estimate the order of the terms, then we
obtain the result $n_0 \propto \bar n^{1-p/2}$. If we use this to
estimate the order of the terms, and omit all terms on the right-hand
side of order higher than $\bar n^{-(p-1/2)}/n_0$, then 
\erf{first3} simplifies to
\begin{align}
-2&\beta\nu \sum_{n=0}^\infty (n+1)^{-p}\braket{\alpha,\zeta}{n}\braket
{n+1}{\alpha,\zeta} \approx -n_0 \bar n^{-(p+1/2)} \nn \\ & +
\nb^{-(p-1/2)}\left\{1+\frac {1}{2n_0}\left[p(p+1)-\frac 14 \right]\right\}.
\end{align}
The first term on the right-hand side is not of higher order than
$\bar n^{-(p-1/2)}/n_0$ if $p\le 1$. If we were estimating the order from
the parameters optimized for minimum intrinsic phase variance then
the first term would be omitted. Now we can expand $-2\beta\nu$ to
give
\begin{equation}
-2\beta\nu \approx \bar n^{1/2}\left( 1- \frac {1}{4n_0}\right)^{1/2}
\left(1- \frac {n_0}{\bar n} \right).
\end{equation}
If we were estimating the order from the parameters optimized for
minimum intrinsic phase variance, the term $n_0/\bar n$ would be
omitted. This then gives us
\begin{align}
&\left(1-\frac {n_0}{\bar n} \right)\sum_{n=0}^\infty (n+1)^{-p}
\braket{\alpha,\zeta}{n}\braket{n+1}{\alpha,\zeta} \nn \\ & \approx -n_0
\bar n^{-(p+1)}+\nb^{-p}\left\{1+\frac{1}{8n_0}+\frac{1}{2n_0}
\left[p(p+1)-\frac 14 \right]\right\}\! . \nn \\
\end{align}
The two terms that would be omitted if we were considering parameters
optimized for minimum intrinsic phase variance just cancel, giving the
simple result
\begin{equation}
\sum_{n=0}^\infty (n+1)^{-p} \braket{\alpha,\zeta}{n}\braket{n+1}
{\alpha,\zeta} \approx \nb^{-p}\left[ 1+\frac{p(p+1)}{2n_0} \right]\!. \nn
\end{equation}


\begin{thebibliography}{99}
\bibitem{fullquan} H. M. Wiseman and R. B. Killip, Phys. Rev. A {\bf 57}, 2169
(1998).
\bibitem{semiclass} H. M. Wiseman and R. B. Killip, Phys. Rev. A {\bf 56}, 944
(1997).
\bibitem{Wise95} H. M. Wiseman, Phys. Rev. Lett. {\bf 75}, 4587 (1995).
\bibitem{Dav76} E. B. Davies, {\em Quantum Theory of Open Systems}
(Academic Press, London, 1976). 
\bibitem{Hel76} C. W. Helstrom, {\em Quantum Detection and Estimation Theory}
(Academic Press, New York, 1976).
\bibitem{Hol84} A. S. Holevo, in {\em Quantum Probability and Applications to
the Quantum Theory of Irreversible Processes}, edited by L. Accardi, A.
Frigerio, and V. Gorini, Springer Lecture Notes in Mathematics Vol. 1055
(Springer, Berlin, 1984), p. 153.
\bibitem{Opatrny} T. Opatrn\'y, J. Phys. A {\bf 27}, 7201 (1994).
\bibitem{Lon27} F. London, Z. Phys. {\bf 40}, 193 (1927).
\bibitem{LeoVacBohPau95} U. Leonhardt, J. A. Vaccaro, B. B{\"o}hmer, and H.
Paul, Phys. Rev. A {\bf 51}, 84 (1995).
\bibitem{Wise96} H. M. Wiseman, Quantum Semiclassic. Opt. {\bf 8}, 205 (1996).
\bibitem{meanN} G. S. Summy and D. T. Pegg, Opt. Commun. {\bf 77}, 75 (1990).
\bibitem{DArPar94}
G. M. D'Ariano and M. G. A. Paris, Phys. Rev. A {\bf 49}, 3022 (1994).
\bibitem{Collett93}  M. J. Collett, Phys. Scr. {\bf T48}, 124 (1993).
\bibitem{Erdelyi53} {\it Higher Transcendental Functions}, edited by A.
Erd\'{e}lyi (McGraw-Hill, New York, 1953), Vol. 1, Sec. 3.15.1. 
\bibitem{Car93b} H. J. Carmichael, {\em An Open Systems Approach to Quantum
Optics} (Springer-Verlag, Berlin, 1993).
\bibitem{Yuen76}  H. P. Yuen, Phys. Rev. A {\bf 13}, 2226 (1976).
\end{thebibliography}
\end{document}